    \documentstyle[nato,psfig,]{crckapb}

\def\kms{km s$^{-1}$}

\def\M{$M_{\odot}$}

\def\m100{mag/100$^d$}
\def\ni{{$^{56}$Ni}}
\def\c57{{$^{57}$Co}\/}

\def\ti44{{$^{44}$Ti}\/}
\def\r0{{$R_0$}}

\begin{opening}
\title{THE PRESENT RATE OF SUPERNOVAE}

\author{M. TURATTO$^1$}
\institute{Osservatorio Astronomico di Padova\\
           vicolo dell'Osservatorio 5, 35122 Padova, Italia}

\end{opening}

\runningtitle{THE RATE OF SUPERNOVAE}

\begin{document}

\begin{abstract} We present and discuss the most recent determination of the rate of
Supernovae in the local Universe. A comparison with other results shows a general
agreement on the gross values but still significant differences on the values of the
rates of various SN rates in different kinds of galaxies. The rate of SNe, used as a
probe of Star Formation, confirms the young progenitor scenario for SNII+Ib/c. The
increasing diversity of SNe reflects also in the SN yields which may affect the chemical
evolution of the Galaxy but, because of the limited statistics, we cannot
estimate the contributions of the new subtypes yet. 
It is also expected that in a few years observational determinations of the SN rates
at various look-back times will be available.

\end{abstract}

\section{Introduction}

In the context of the chemical evolution of the galaxies the supernova (SN) rate and its
history play a fundamental role. But, while there are several theoretical models
predicting the time evolution of the SN rates with time, observational determinations at
different look-back times are not yet available. An estimate of the rate of SNIa at
z$\sim0.4$  has only be published so far (Pain et al. 1997). The efforts of the high-z
searches have been focused more on the detection and follow--up of the SNe for the
determination of the geometry of the Universe than on the careful analysis of their
databases for the SN rates.

Different is the case of the estimates of the rate in the local Universe. Searches of
bright SNe are active since several decades and their archives have been
exploited to this purpose since the sixties.
The determinations of the rates have been carried out on different samples and 
with different techniques. In some case the differences among various estimates
are larger than 100 \%. In this paper we discuss the reasons of these 
uncertainties and their relation to the chemical evolution of the Galaxy.

\section{Definition of SN rate}

The SN rate is observationally defined in its general form as the number of SNe
exploding in a given sample of galaxies during a certain 
time interval

$$ \nu(yr^{-1}) = N_{SN} N_{gal}^{-1} \Delta t^{-1} $$

and is measured in yr$^{-1}$. However, since the galaxies can have very
different stellar content, one expects that the average value depends strongly 
on the composition of the galaxy sample.
Indeed, with the growing number of discoveries it appeared that critical
parameters for the SN rates were the galaxy type, with late spirals more
prolific than early spirals, and the galaxy luminosity, which in first
approximation measures the number of available progenitor stars
(Tammann 1970, Cappellaro \& Turatto 1988).

The SN rate can therefore be computed for each galaxy type as

$$ \nu^{gtype}(SNu) = N_{SN}^{gtype} (\sum_j^{Ngal} \sum_i^{obs} k_i L_j^{gtype}
\Delta t_i^{gtype})^{-1} $$

where the index $j$ refers to the sample galaxies of that type and $i$ refers
to the individual observations of each  galaxy. Convenient units for the rate 
are the SNu, where 1 SNu  corresponds to 1 SN per $10^{10}$ solar 
luminosities per century. Because of the
normalization to galaxy luminosity, SNu scale as H$_o^2$. The parameter $k_i$
accounts for various observational selection effects.

\section{Efforts for better SN rates}

It is evident from the last equation that in order to have better estimates of
the SN rates one can: 1) enlarge the statistical samples (N$_{SN}$, N$_{gal}$),
2) estimate with higher precision the surveillance times, or 3) determine more
reliable correction terms for the selection effects.

For the sample selection there are in practice two alternatives. 
We can, for instance, select all
galaxies within a given volume, making reasonable assumptions on the
surveillance times (cfr. Tammann 1994). Consequently the SN rate is computed
using all the SNe discovered inside the volume during that period. The main
advantage of this approach is in the large samples directly available form the
galaxy and SN catalogues (e.g. Barbon et al. 1999). The drawback 
is in the critical assumption of the surveillance time which can hardly be the
same for each galaxy.

The alternative is the survey approach which limits the galaxies to those
surveyed by a specific SN search and the SNe to those discovered during the same 
search. The method allows for each galaxy a precise determination of the {\it
control time}, i.e. the time during which a possible SN of a given type remains
above the detection limit. This determination takes into account the
distribution of the observations during the survey, the distance of the galaxy, the
properties of the SN types and the observational limitations of the survey. The
method, introduced already by Zwicky (1938), found several
applications (e.g. Cappellaro \& Turatto 1988, Evans et al. 1989). In this case the
drawback is the small size of the sample.

In order to overcome such limitation we developed a strategy to treat
simultaneously the databases of several of SN searches (Cappellaro et al. 1993, 1997a).
The use of the five SN searches listed in Tab.~\ref{searches} led us (Cappellaro et al.
1999) to build  the largest sample ever used for the determination of SN rates, 137 SNe
in 9346 RC3 galaxies, for a total control time of the sample of 24954 yr.

\begin{table}
\begin{center}
\caption{The pooled SN searches} \label{searches}
\begin{tabular}{llll}
\hline 
SN search & type($^*$) & SN discoveries($^{**}$) & reference\\
\hline
Asiago	      & pg	& 51 & Cappellaro et al. 1993  A\&A 268, 472\\
Crimea 	      & pg	& 33 & Tsvetkov 1983 SA 27, 22\\
O.C.A.	      & pg	& 16 & Pollas 1994, in Les Houches 1990, Bludman et al. p.769 \\  
Calan/Tololo  & pg	& 12 & Hamuy et al. 1993 AJ 106,  2392\\
Evans	      & vis	& 54 & Evans 1997 PASA 14, 204\\
\hline
\end{tabular}

($^*$) pg=photographic; vis=visual\\
($^{**}$) some of which earlier reported by other searchers
\end{center}
\end{table}

The determination of the correction terms for the biases in the discovery of SNe
is crucial and controversial. Two biases are by far the most important: the
effect of inclination in spirals and the deficiency of discoveries in the
central regions of distant galaxies.

Several authors claimed that the bias due inclination was not affecting visual and CCD
searches but only photographic ones (Evans et al. 1989, Muller et al. 1992). The
composite sample above allows us to perform a direct check handling separately
photographic and the Evans' visual searches. The rate of SNe computed for the Evans'
sample with the control--time technique  described above is 2.6 times greater in
face--on spirals than in inclined ones, showing that this bias is almost as severe as in
photographic searches for which the factor is 2.9. The fact that different detectors
provide similar results, is an indication that the effect is intrinsic to the galaxies
rather than to the observational technique.

The natural explanation is that SNe in inclined spirals suffer in average
heavier dust extinction than those in face--on galaxies, thus are more difficult
to discover. If this interpretation is correct we expect (1) that searches
working at longer wavelengths are less affected, (2) that the effect is stronger
in late type spirals and (3) that it is more important for core--collapse SNe
which are associated to a younger population, than for SNIa associated to an
older population.

So far contradictory answers have been given to the points above. The effect
seems stronger in late than in early type galaxies (Cappellaro et al. 1997a),
only marginally smaller in the visual that in the blue bands, as said above,
(probably significantly smaller for red CCD searches) but comparable for
core--collapse and thermonuclear SNe (Cappellaro et al. 1997a). Larger
statistics is necessary for unambiguous answers.

In the attempt to correct for the bias due to inclination we have tried two different
approaches (Cappellaro et al. 1999). First we applied the $\sec\alpha$ term
corresponding to a plane parallel distribution of the dust in the disks of spirals. This
results in an apparent overcorrection of the SN rate in edge--on galaxies. Then we
explored the possibility to apply a model for the dust and the SN distributions
consistent with our knowledge of the galaxies and SN progenitors. Even this more
physical treatment showed some limitations indicating that the dust and the progenitor
distributions in real galaxies are more complex than in simple models.

The study of the deficiency of SN discovery in the central regions of the
parent galaxies has been discussed in detail in Cappellaro \& Turatto (1997).
The effect is critical in distant galaxies with about 50\% of SNe exploding in
galaxies with recession velocities larger than 7500 \kms\/ lost in the central
parts, in particular, the innermost 10 arcsecs. Interestingly the effect seems
correlated to the inclination in spirals with more inclined galaxies more
affected. The effect for the SN searches of Tab.\ref{searches} was empirically
determined in Cappellaro et al. (1997a) and accounted for also in Cappellaro et
al. (1999).

\section{The SN rates per unit B luminosity}

The SN rates derived with the procedure and the sample described above 
are reported in Table~\ref{rates}. The errors include those
due to the statistics and those due to the input parameters and the 
selection effects.

\begin{figure}
\begin{center}
\psfig{figure=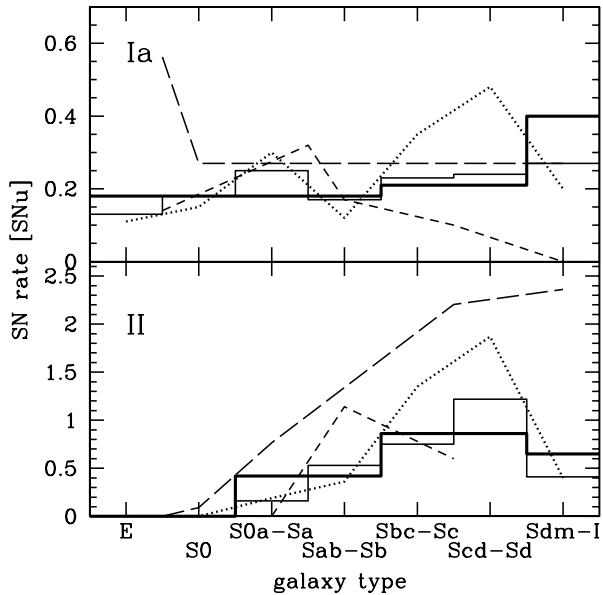,height=3.5in}
\caption{Comparison between different recent estimates of the SN rates scaled
to H$-o=75$ \kms\/ Mpc$^{-1}$
(Cappellaro et al. 1999, thick continuos line;
Cappellaro et al. 1997a, thin continuous line; 
Tammann et al. 1994, long dashed line;
van den Bergh \& McClure 1994, short dashed line;
Cappellaro et al. 1993, dotted line).}
\label{cfr}
\end{center}
\end{figure}

A comparison of these rates with other recent determinations is shown in
Figure~\ref{cfr}. There is a reassuring overall  agreement among rates based on
different samples and with different recipes. The largest differences are with Tammann
et al. (1994) and for SNIa in ellipticals and for SNII. We believe that such differences
are mainly due to biases in the SN discoveries in the fiducial sample. Our new rates of
SNIa show a shallower dependence on the galaxy types than our previous works of 1997 and
1993, which were based on the same approach but on smaller samples and with different
bias corrections. At that time the increase of the rate of SNIa per unit luminosity
moving from early to late type galaxies was considered as an argument in favor of
different ages of the precursors of SNIa, with  progenitors in spirals younger than
those in ellipticals. This new result might be considered in favor of the conventional
scenario of similarly old progenitors in all galaxy types. We warn that the high rate of
SNIa observed in the heterogeneous class of Sm+Irregulars+Peculiars is uncertain to
$\pm40$\%.

As in the previous works, the rates of SNII increase from early to late spirals
but now the peak values are smaller. This is mainly due to the smaller
correction terms used for the inclination effect. Similar trends are shown also
by SNIb/c which however remain definitely less frequent than type Ia.

\begin{table}
\begin{center}
\caption{SN rate(in SNu(B)) for the combined sample.}\label{rates}
\begin{tabular}{ccccc}
\hline  
galaxy       &\multicolumn{3}{c}{rate [SNu]} \\
type         &  Ia   &   Ib/c &  II    & All\\
\hline
E-S0         & $0.18\pm0.06$ & $<0.01$       & $<0.02$       & $0.18\pm0.06$\\
S0a-Sb       & $0.18\pm0.07$ & $0.11\pm0.06$ & $0.42\pm0.19$ & $0.72\pm0.21$\\
Sbc-Sd       & $0.21\pm0.08$ & $0.14\pm0.07$ & $0.86\pm0.35$ & $1.21\pm0.37$\\
Sm,Irr,Pec   & $0.40\pm0.16$ & $0.22\pm0.16$ & $0.65\pm0.39$ & $1.26\pm0.45$\\
\hline
\end{tabular}

$1 \,{\rm SNu(B)} = 1\, {\rm SN}\, (100 {\rm yr})^{-1}\, (10^{10}
L_\odot^{\rm B})^{-1}$, H$_o=75$ \kms\/ Mpc$^{-1}$\\
\end{center}
\end{table}

\section{SN rates vs. Star Formation Rate} 

The normalization of the rates to the B luminosity of the galaxies has been a
natural choice. In fact, the B magnitude is the magnitude most frequently
available in the literature and it scales, in first approximation, with the
galaxy mass since stars of different population have significant emission in
this band. 

In principle, the normalization of the rates to photometric bands sensitive to
specific stellar populations might give important insights onto the SN
progenitor scenario.  If in all galaxies the flux at a given wavelength were
due to a single stellar population, then the rate of the SNe associated to this
population, normalized to the luminosity in this band, should be constant. For
instance, infrared H or K bands would be particularly suitable for the study of
the type Ia progenitors since they sample well the old stellar population.
Unfortunately, the  sample of galaxies with published infrared photometry is
still too small for any direct analysis. Different is the situation for the
integrated far infrared luminosities which are available for a large sample of
galaxies thanks to the IRAS survey. L$_{FIR}$ has been used often as an
indicator of SFR and therefore we may try to gain information on
the progenitors of core--collapse SNe.

Contrary to the expectations, the rate of core--collapse SNe (cfr. column 3 of
Table~\ref{lfir}) varies of almost a factor two moving from early and late
spirals. Taken to face value this means that SNII and Ib/c are not associated
to massive progenitors. Actually we think that this inconsistency is due
to the contamination of the FIR emission associated with dust around massive
stars by the cool component associated to the general stellar radiation field.
In other words we consider this a further evidence that in normal galaxies
L$_{FIR}$ is not a reliable probe of SFR.

This finding is supported by the following argument. The effect of the general
radiation field, due also to old stars, can be partially removed normalizing
the FIR to the B luminosity to get the so--called infrared excess. The SN
rates (in SNu(B)) computed independently for the three classes of IRAS not
detected, low and high infrared excess galaxies are similar for type Ia.
Instead for type II+Ib/c SNe the rate of IRAS not detected galaxies is
significantly lower than in galaxies with high FIR excess (Table~6 of
Cappellaro et al. 1999) in agreement with the young progenitor scenario.

\begin{table}
\begin{center}
\caption{SN rates per unit FIR luminosity.  }
\label{lfir}
\begin{tabular}{lcccc}
\hline  
galaxy  &\multicolumn{3}{c}{SN rate [SNuIR]} \\
type    &  Ia   &   II+Ib/c    & All\\
\hline 
E-S0    & $1.8\pm0.8$ &             & $1.8\pm0.8$\\
S0a-Sb  & $0.6\pm0.2$ & $2.0\pm0.5$ & $2.7\pm0.5$\\
Sbc-Sd  & $0.6\pm0.1$ & $3.5\pm0.6$ & $4.1\pm0.6$\\
\hline   
\end{tabular}

$ 1\,{\rm SNuIR} = 1\, {\rm SN}
(100 yr)^{-1} (10^{10} L_{{\rm FIR},\odot})^{-1}. $\\
\end{center}
\end{table}

Also the colors of galaxies can be used as indicators of the stellar populations with
bluer galaxies hosting younger, more massive stars. We computed the SN rates (in SNu(B))
in galaxies with different optical integrated colors (Cappellaro et al. 1999). We found
that the observed rate of SNIa is independent on the galaxy colors while the rate of
core--collapse SNe is higher in bluer spirals than in galaxies of the same morphological
type but with redder colors.

It is also interesting to note that the number of massive progenitors predicted by
the star formation rate, derived by evolutionary synthesis models of galaxies
with different colors, closely matches the observed number of core--collapse
SNe.

\section{Frequency and  Nucleosynthesis}

Supernovae are primary contributors to the chemical enrichment of the
galaxies because they return to the ISM the elements synthesized by the
hydrostatic and explosive burning processes, with the exception of the
material trapped in the condensed remnants of core--collapse SNe.
From a general point of view core--collapse SNe, descending from
young massive progenitors determine the enrichment in the early galactic
evolution while supernovae of type Ia are responsible of the heavy elements
enrichment in the later phases of the galactic evolution on time scales which
depend on the binary configuration and mass ratios.
Current efforts are devoted to determine the masses of enriched
material released by different kinds of explosions given their relative frequency.

For type II it is expected that the yields vary with the stellar mass, mass cut,
explosion energy and neutron excess (Nakamura et al. 1999). The effect of these
parameters have been modeled theoretically but observational confirmations are still
needed. The well studied explosions of SN~1987A, SN~1993J and SN~1994I with progenitor
masses between 13 and 20 \M\/ produced about 0.07 \M\/ of \ni. Early works on the late
light curves indicated that most SNII produced similar amounts of radioactive material
(Turatto et al. 1990). In the last years data have been accumulated which show a
different scenario (Danziger, this meeting). In particular there seems to be an increase
of the \ni\/ production for stars up to about 25\M\/ (e.g. SN~1997ef, 0.15\M\/ of \ni,
and 1998bw, 0.7 \M) after which the gravitational potential wins and a small \ni\/ mass
is ejected due to fallback (e.g. SN~1997D, 0.002\M of \ni\/; Turatto et al. 1998). The
variation of the mass cut as a function of the progenitor mass seems to account for the
observed trends of iron peak element abundance ratios in stars of low metallicity
(Nakamura et al. 1999). The study of a second object with an expected low \ni\/
production, SN~1999eu a twin of SN~1997D at early time, will help to strengthen this
hypothesis. Unfortunately the objects for which both the progenitor mass and the ejected
\ni\/ mass are available is  so limited that no kind of statistics is still possible.

Also for type Ia SNe, once considered a very homogeneous class, there is now evidence
that the yields can be different. Cappellaro et al. (1997b) showed that the photometric
properties of SNIa require a range of a factor 10 of the radioactive material
synthesized during the explosion (0.1 to 1 \M) and a factor 2 in the total mass of the
ejecta. From a theoretical point of view such variations can be accounted for with
different precursor masses (Chandrasekhar and sub-Chandrasekhar), and flame speeds. 
An attempt to
determine the relative frequency of different Type Ia subtypes has been performed by
Cappellaro et al. (1997a). Again going to analyze very particular SN subtypes we faced
the problem of poor statistics. For faint SNIa, producing small amounts of \ni, we found
an intrinsic rate about 1/4 of that of all SNIa although they constitute only 5\% of all
SNIa discoveries. Also the percentage of SNIa producing large amounts of \ni\/ is
expected to be small. In fact the observed number is relatively small although they are
brighter than the average, hence easier to discover.

\section{Conclusions}

The  determinations by various authors of the rates of SNe in the local Universe
are in substantial agreement although large differences are still present on the
rates of individual SN types in specific galaxy types. Improvements are possible
in the treatment of the databases and in the analyses of the selection effects.
However the slow increase of the statistics at low redshifts is the limiting factor.

On the contrary, huge is the number of discoveries at high (z$=0.5-1.0$) redshifts
thanks to the projects aiming to determine the geometry of the Universe with the use of
SNe as distance indicators. The analysis of these databases can already  allow the
determination of the rates at such early epochs. Data at intermediate redshifts are
still missing. In order to cover this gap we have started a SN search in the southern
hemisphere which is starting to produce the first interesting results (SNe 1999ey,
1999gt, 1999gu). In few years observational estimates of the SN rates at various
look-back times will provide strong constraints to the Galaxy evolutionary models.


\begin{thebibliography}{}
\bibitem[]{} Barbon,R., Buond\'i, V., Cappellaro, E., Turatto, M., 1999,
A\& AS 139, 531
\bibitem[]{} Cappellaro, E., Turatto, M., 1988, A\&A 190, 10
\bibitem[]{} Cappellaro, E., Turatto, M. 1997, in:  Nato -- ASI on
   Thermonuclear Supernovae, eds. R. Canal, P. Ruiz-Lapuente,
   J. Isern, Kluwer Academic Publisher, Dordrecht p. 77
\bibitem[]{} Cappellaro, E., Turatto, Benetti,S., Tsvetkov, D.Yu., 
   Bartunov, O.S., Makarova, I.N., 1993, A\&A 273, 383
\bibitem[]{} Cappellaro, E., Turatto, M., Tsvetkov, D.Yu., Bartunov, O.S.,
   Pollas, C., Evans, R., Hamuy, M., 1997a, A\&A 322, 431
\bibitem[]{} Cappellaro, E., Mazzali,P., Benetti,S., Danziger, I.J.,
   Turatto, M., Della Valle,M., Patat,F.,  1997b, A\&A 328, 203

\bibitem[]{} Cappellaro, E., Evans,R., Turatto, M., 1999, A\&A 351, 459
\bibitem[]{} Evans,R., van den Bergh,S., McClure,R.D., 1989, ApJ 345, 752
\bibitem[]{} Muller, R.A., Marvin, H.J., Pennypacker, C.R., Perlmutter, S.,
   Sasseen, T.P., Smith, C.K. 1992, ApJ 384, L9 
\bibitem[]{} Nakamura,T., Umeda,H., Nomoto,K., Thielemann,F., Burrows,A.,
   1999, ApJ 517, 193
\bibitem[]{} Pain, R., et al. 1996, ApJ, 473, 356 
\bibitem[]{} Tammann, G.A., 1970, A\&A 8, 458
\bibitem[]{} Tammann, G.A., L\"offler,W., Schr\"oder,A., 1994, ApJS 92, 487
\bibitem[]{} Turatto,M., Cappellaro,E., Barbon,R., Della Valle,M., Ortolani,S.,
   Rosino,L., 1990, AJ 100, 771
\bibitem[]{} Turatto,M., et al. 1998, ApJL 498, 129
\bibitem[]{} van den Bergh, S., McClure, R.D. 1994, ApJ 425, 205
\bibitem[]{} Zwicky,F., 1938, ApJ 96, 28

\end{thebibliography}
\end{document}